\documentclass[final,1p,times,authoryear]{elsarticle}
\usepackage{amssymb}
\usepackage{lineno}

\journal{The Cell Surface}

\begin{document}

\begin{frontmatter}

%% Title, authors and addresses

%% use the tnoteref command within \title for footnotes;
%% use the tnotetext command for theassociated footnote;
%% use the fnref command within \author or \address for footnotes;
%% use the fntext command for theassociated footnote;
%% use the corref command within \author for corresponding author footnotes;
%% use the cortext command for theassociated footnote;
%% use the ead command for the email address,
%% and the form \ead[url] for the home page:
%% \title{Title\tnoteref{label1}}
%% \tnotetext[label1]{}
%% \author{Name\corref{cor1}\fnref{label2}}
%% \ead{email address}
%% \ead[url]{home page}
%% \fntext[label2]{}
%% \cortext[cor1]{}
%% \address{Address\fnref{label3}}
%% \fntext[label3]{}

\title{Elongation and shape changes in organisms with cell walls: a dialogue between experiments and models}

%% use optional labels to link authors explicitly to addresses:
%% \author[label1,label2]{}
%% \address[label1]{}
%% \address[label2]{}

\author[label1,label2]{Jean-Daniel Julien}
\author[label2]{Arezki Boudaoud\corref{cor1}}
\address[label1]{Laboratoire Reproduction et D\'eveloppement des Plantes, Universit\'e de Lyon, ENS de Lyon, UCB Lyon 1, CNRS, INRA, 46 all\'ee d'Italie, 69364 Lyon Cedex 07, France}
\address[label2]{Laboratoire de Physique, Univ. Lyon, ENS de Lyon, UCB Lyon 1, CNRS, 46 all\'ee d'Italie, 69364 Lyon Cedex 07, France}
\cortext[cor1]{\ead{arezki.boudaoud@ens-lyon.fr}}

\begin{abstract}
%% Text of abstract
The generation of anisotropic shapes occurs during morphogenesis of almost all organisms. With the recent renewal of the interest in mechanical aspects of morphogenesis, it has become clear that mechanics contributes to anisotropic forms in a subtle interaction with various molecular actors. Here, we consider plants, fungi, oomycetes, and bacteria, and we review the mechanisms by which elongated shapes are generated and maintained. We focus on theoretical models of the interplay between growth and mechanics, in relation with experimental data, and discuss how models may help us improve our understanding of the underlying biological mechanisms.
\end{abstract}

\begin{keyword}
%% keywords here, in the form: keyword \sep keyword
morphogenesis \sep cell wall \sep bacteria \sep fungi \sep yeasts \sep oomycetes \sep plants \sep symmetry breaking \sep cell polarity

%% PACS codes here, in the form: \PACS code \sep code

%% MSC codes here, in the form: \MSC code \sep code
%% or \MSC[2008] code \sep code (2000 is the default)

\end{keyword}

\end{frontmatter}

 % \linenumbers

\section{Introduction}

Symmetry breaking is a fascinating feature of morphogenesis: How does a sphere-like organism become rod-like? How does a rod-like organism maintain its shape? Symmetry breaking occurs within cells, when cell polarity is established and proteins or organelles are asymetrically distributed in the cytoplasm or at the plasma membrane~\citep{bornens2008,goehring_cell_2013}. Cell polarity has been extensively investigated, notably using modelling approaches~\citep{mogilner2012}. In this review, we consider symmetry breaking when it is associated with shape changes, as exemplified by branching during hydra development~\citep{mercker_mechanochemical_2015}. More generally, symmetry breaking occurs when an initially symmetric shape --- e.g. a sphere that is unchanged by rotations around its centre, or a cylinder that is unchanged by rotations around its axis --- or an initially symmetric distribution of molecules (e.g. homogeneously distributed on a sphere) becomes asymmetric --- a bump appears on the sphere or on the cylinder, or the molecules become more concentrated around a point of the sphere. Here we focus on walled cells and particularly consider plants, fungi, oomycetes, and bacteria. The cells of these organisms are characterised by a relatively high internal hydrostatic pressure known as turgor pressure, which results from the concentration of solutes in the cytosol, and drives growth \citep[though this is debated, see e.g.][]{harold1996, rojas_response_2014}. Turgor pressure is counterbalanced by a rigid shell, the extra-cellular matrix known as the cell wall, that prevents cells from bursting~\citep{harold_force_2002}. As such, the cell wall is fundamental in determining cell shape. Although the cell walls of these organisms differ in their chemical composition~\citep{lipke_cell_1998,cosgrove_growth_2005,silhavy_bacterial_2010}, they have similar mechanical properties that are regulated so as to shape cells. In particular, pressure being a global and isotropic (non-directional) force, inhomogeneous or anisotropic distributions of mechanical or biochemical properties seem needed to establish and maintain asymmetric shapes. 

Previous reviews addressed such questions for specific systems, see for instance \citep[][for fission yeast]{davi_mechanics_2015}, \citep[][for bacteria and fission yeast]{chang_how_2014}, \citep[][for pollen tubes]{kroeger_pollen_2012}, or \citep[][for plants]{uyttewaal_integrating_2010}. In this review, we discuss the links between mechanics and growth of elongated cells across kingdoms. We pay specific attention to computational modelling, because it appears as a powerful tool to validate hypotheses by setting aside all but the fundamental actors of the phenomena investigated and to guide future experimental effort. We apologise to those whose work could not be included.

\section{Systems of interest and their cell walls}
We first  present a few model systems; we give a brief introduction to their cell walls in terms of composition, structure, and mechanical properties, along with a presentation of available mechanical models of cell walls.

\subsection{Rod-shaped bacteria}
All kinds of shapes are found amongst bacteria. Actually, shape has long been an important criterion for classification~\citep{cabeen_bacterial_2005} and is fundamental for many bacterial functions~\citep{young_selective_2006,singh_determining_2011,chang_how_2014}. The rod is a very common shape, with \emph{Escherichia coli} as a representative Gram-negative species, making bacteria good systems to study the establishment and maintenance of elongated shapes. Many rod-shaped bacteria expand diffusively, with new cell wall incorporated all along the rod, while other expand at a pole or within a restricted region~\cite{cava2013}.

The typical bacterial cell wall is mostly made of peptides and glycans~\citep{cabeen_skin_2007,silhavy_bacterial_2010}. The glycan strands are oriented circumferentially and give the cell wall anisotropic mechanical properties~\citep{chang_how_2014}, the cell wall being stiffer in the circumferential direction than in the longitudinal direction. In Gram-positive bacteria, the structure of the cell wall is not as well characterised as in Gram-negative bacteria, although recent studies also support a circumferential arrangement of glycan strands~\citep{beeby_architecture_2013}. A detailed mechanical model of the cell wall of Gram-negative bacteria was proposed by~\citet{huang_cell_2008}, with the peptidoglycan network simulated like a network of springs; by adding defects in the network and modulating the type of defects and their density, this model was able to reproduce several bacterial cell morphologies: curved, helical, snake-like, and lemon shapes. Whereas this model did not take growth into account, it was the starting point for many models of wall expansion, which are discussed in the next section.

\begin{figure}
\begin{centering}
\includegraphics[scale=0.3]{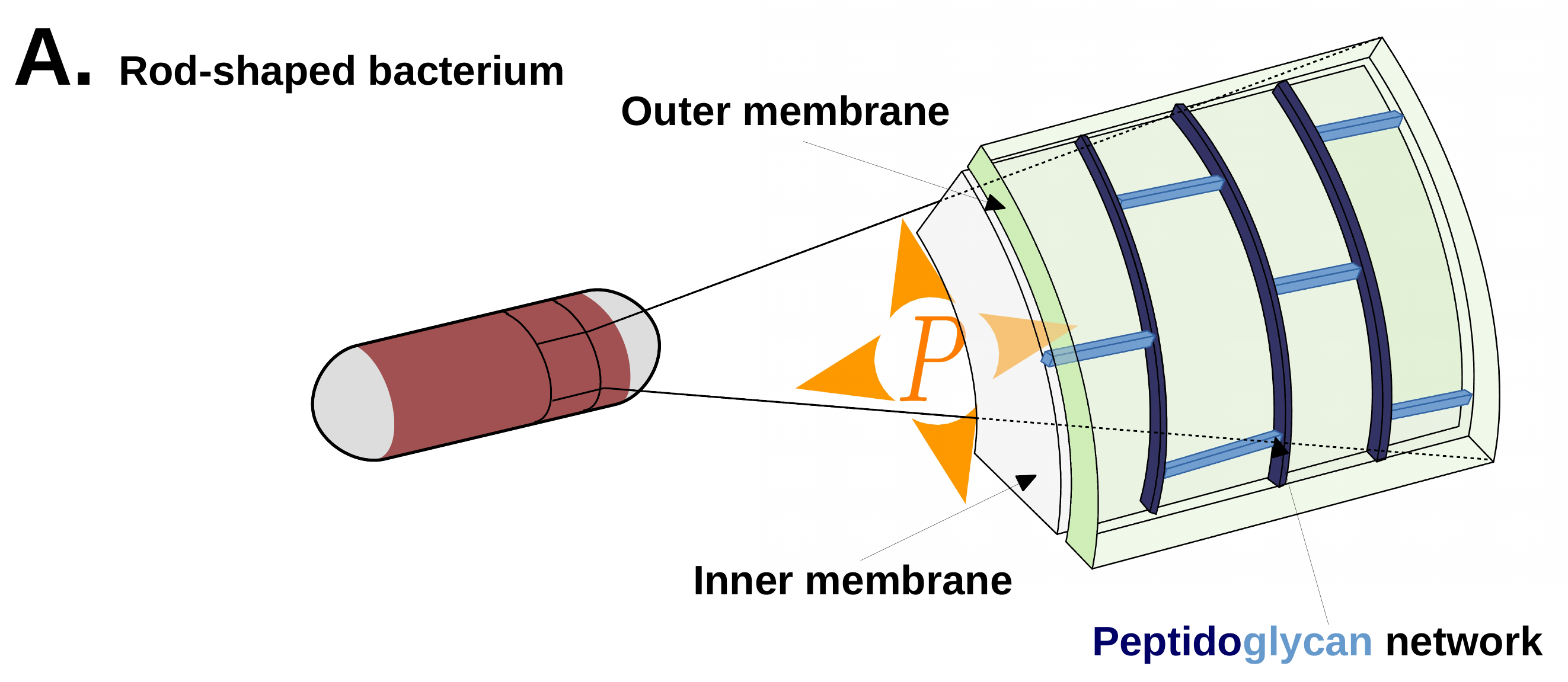}
\par\end{centering}

\begin{centering}
\includegraphics[scale=0.3]{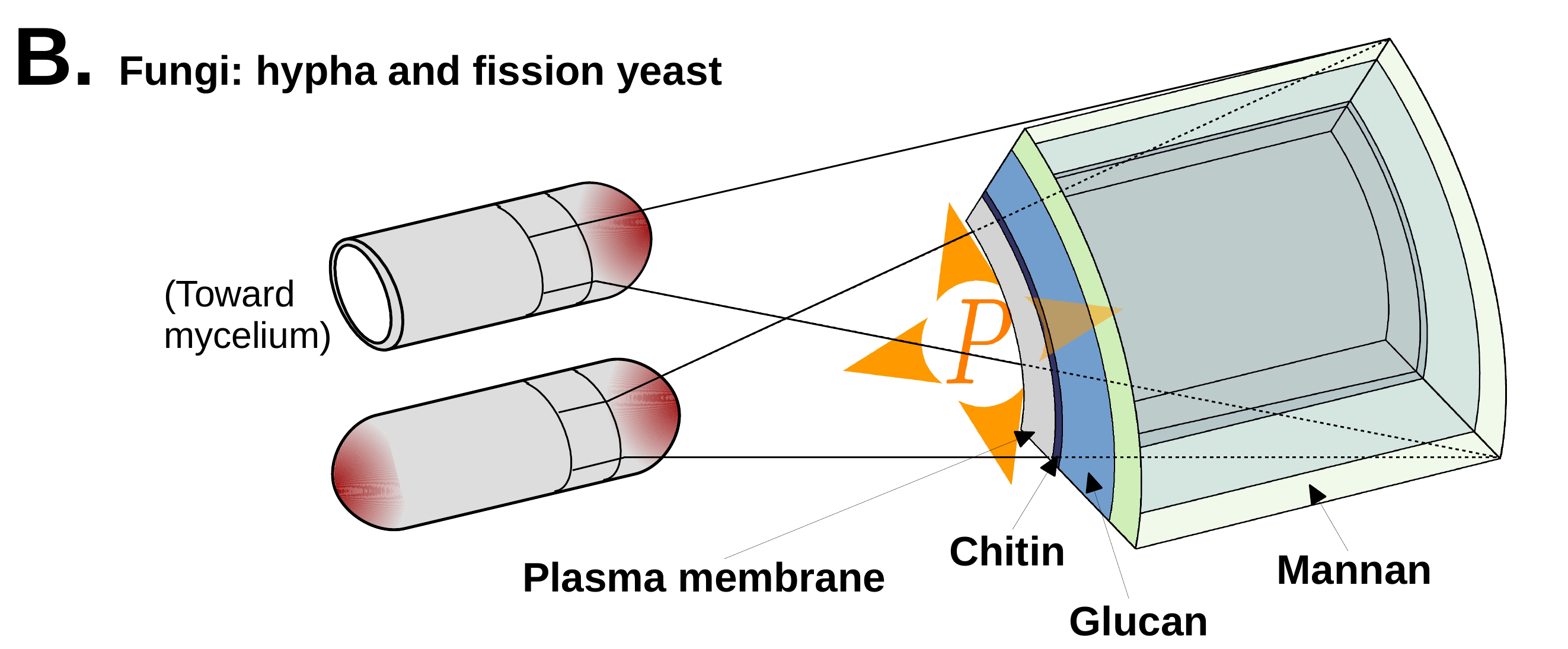}
\par\end{centering}

\begin{centering}
\includegraphics[scale=0.3]{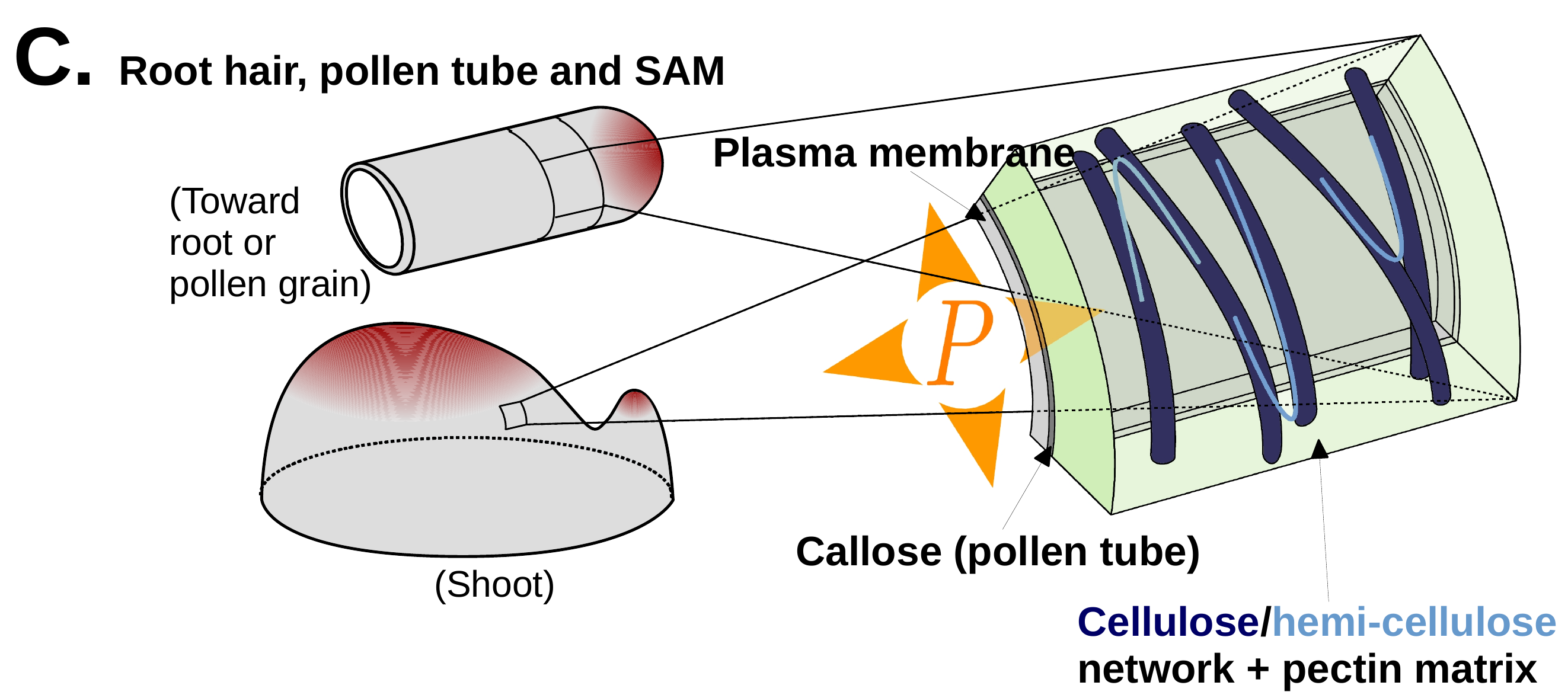}
\par\end{centering}

\caption{\textbf{Systems of interest: growth mode and composition of the cell wall.} (A) In many rod-shaped bacteria, such as \textit{E. coli}, growth is diffuse, localised on the whole cylindrical part on the cell. The cell wall of \textit{E. coli} and other Gram-negative bacteria is composed of a stiff layer of peptides and glycans surrounded by
two lipid membranes \citep{silhavy_bacterial_2010,chang_how_2014}. (B) Tip growth is observed in fungal hyphae and fission yeast. According to models, the cell wall would be composed of three layers, made respectively of chitin, glucan and mannan~\citep{lipke_cell_1998,bowman_structure_2006}. (C) In plants, pollen tubes and root hairs are tip-growing cells. On a larger scale, growth is focused on the tip of emerging organs around the shoot apical meristem. The plant cell wall is made of a network
of cellulose embedded in a matrix of hemicellulose and pectin \citep{cosgrove_growth_2005}. Callose can also be found, notably in pollen tubes \citep{chebli_gravity_2011}. In the three systems, the relative amounts of components may vary, for instance between species or even spatially within a single cell.}
\end{figure}

\subsection{Fungi and oomycetes}

Many fungi have elongated cells that grow from their tips: hyphae and yeasts. Hyphae are very long filamentous cells, that can be collectively organised into a mycelium. Yeasts are unicellular fungi. In particular, fission yeast (\textit{Schizosaccharomyces pombe}) is a good system to study cell polarity and the maintenance of rod shapes~\citep{chang_shaping_2009}. Indeed, it grows as a capped cylinder, maintaining a constant diameter (except for spores, which are roughly spherical).

The cell wall of fungi is mostly made of glucans (excluding cellulose), mannoproteins, and chitin~\citep{lipke_cell_1998}. Although many of these components are fibrous, it is believed that the fungal cell wall does not have anisotropic mechanical properties because of the lack of preferential orientation of the fibres~\citep{chang_how_2014}. (Strictly speaking, the fibres are mostly tangential, and the cell wall is transversely isotropic, being softer across the thickness than in the directions tangential to the wall). Despite a rather well-known composition, fungal cell walls have not been modelled in detail.

Although oomycetes grow in mycelial forms like fungi, they belong to a different taxonomic group, Stramenopiles. Their cell walls are mostly made of glucans and, unlike fungi, they contain some cellulose and tiny amounts of chitin~\citep{melida2013}.

\subsection{Plants: pollen tubes and root hairs}
Plants provide two other systems of interest,  pollen tubes and  root hairs, that elongate through tip growth. The pollen tube is a long protuberance that grows out from the pollen grain until it reaches the ovule for fertilisation~\citep{geitmann_how_2010,kroeger_pollen_2012}. Accordingly, its growth is fast and highly directional. Its growing tip is formed of a single cell. Root hairs are long tubular outgrowths from specialised epidermal cells of the root. They are important for absorption of nutrients and anchorage to the soil~\citep{carol_building_2002,grierson_root_2014}.

The plant cell wall is mostly made of polysaccharides: cellulose microfibrils embedded in a matrix of hemicelluloses and pectins~\citep{varner_plant_1989}. Cellulose fibrils can be much longer than the cell diameter and their organisation differs according to cell type and to developmental stage, ranging from highly directional circumferential alignment to random orientations \citep{cosgrove_growth_2005}. As cellulose is the stiffest component of the wall, a preferential orientation of microfibrils can give anisotropic properties to the plant cell wall: It is stiffer in the direction of the fibres~\citep{kerstens_cell_2001}, which may lead to less expansion in this direction, and so drive anisotropic cell growth. In pollen tubes and root hairs, cellulose usually displays a helical arrangement that could help resist bending forces and penetrate external medium \citep{Aouar2010}; this arrangement could also reinforce the transition region between the tip and the cylindrical part, which bears the highest tension \citep{geitmann_how_2010}. An early model of the cell wall focused on the self-organisation of cellulose due to cell geometry~\citep{emons_making_1998,mulder_dynamical_2001}; the condition of optimal packing of cellulose microfibrils restrains their direction and the movement of their synthesising complexes along the axis of the cell can generate various types of organisations with locally aligned fibres. More recently, models attempted to define realistic geometries for the arrangement of polysaccharides in the cell wall and to predict the corresponding elastic properties for small deformations~\citep{qian_finite_2010,kha_wallgen_2010,yi_architecture_2012}.

\subsection{Plants: A multicellular case, the shoot apical meristem}

Some of the relevant results that we will discuss were obtained in a multicellular context. The shoot apical meristem (SAM) is the tissue located at the tip of any above-ground branch in a plant; it contains a stem cell pool and is the site of organogenesis~\citep{ha_shoot_2010,murray_systems_2012,gaillochet_o_2014}. Organs are initiated around the tip and emerge as protuberances from the apical dome, breaking the symmetry around the axis of the dome. As discussed above, the deposition of cellulose microfibrils in a preferential direction can provide anisotropic mechanical properties to the SAM.

\section{Generating elongated shapes}

\subsection{Tip growth}

We first consider geometrical models built in the context of hyphal growth. In hyphae of many fungi, a intriguing structure localised close to the tip and known as the spitzenk\"orper (SPK) concentrates vesicles and is thought to be the organising centre of tip growth. It has been proposed that vesicles, containing notably multiple cell-wall regulating enzymes, are transported to the region of the SPK by cytoplasmic microtubules. Actin microfilaments, found in the SPK, then take over in regulating the supply of enzymes and material to the membrane. The complete composition of the SPK is still unclear \citep[see][for a review on hyphal growth]{heitman_cell_2017}. Its description as a cluster of cell wall-building enzymes has led to the Vesicle Supply Centre (VSC) model~\citep{bartnicki-garcia_computer_1989} of fungal growth, first implemented in two dimensions. In this framework, the VSC is a point in space that constantly emits vesicles in random directions. Those vesicles move at constant velocity, and locally increase the length of the cell wall after they have reached it. Finally, the VSC moves at a constant, prescribed velocity. This yields a steady shape that compares well with experimental observations of growing hyphae. The three-dimensional generalisation of the VSC model~\citep{gierz_three-dimensional_2001} raised the question of how the material brought by the vesicle is distributed between the longitudinal and the circumferential directions, which led to propose additional rules for expansion.  A second improvement of the VSC model was to replace the ballistic motion of the vesicles by diffusion \citep{tindemans_diffusive_2006}, which only slightly modifies the shapes generated by the model. VSC models were successful in demonstrating that self-similar tip growth can emerge from patterns of exocytosis. However, they assumed that wall expansion is limited by supply of materials, and that cell wall mechanics is negligible. While there is experimental evidence that exocytosis is required for growth, it is also clear that cell wall mechanics is important to set the pace of expansion~\citep{kroeger_pollen_2012}. 

Cell wall mechanics was accounted for in several generic models. A first class of models assumes that growth can be considered as a viscous process, whereby the cell wall expands like a viscous material under the tension generated by turgor pressure; such a process would lead to wall thinning, and so cell wall synthesis is assumed to maintain the cell wall thickness at an approximately constant value. \citet{bernal_mechanics_2007} got inspiration from the inflation of rubber balloons, which were modelled as elastic shells with spatially varying stiffness: More precisely, the compliance of the material (how easy to stretch it is) is large on a narrow annular region around the tip and small on the cylindrical part of the shell. This model was able to reproduce observed deformations of root hairs. This work highlighted the importance of modulating the global pressure drive by local supply of cell wall material, or by local modifications of cell wall properties. The spatial extent of the wall deposition, and more precisely how it depends on the size of the cell, was theoretically found to change the shape of the growing tip~\citep{campas_shape_2009}. In this latter study, the cell wall was modelled as a thin viscous shell with infinite viscosity on the flanks (so that the tube maintains its diameter). Growth is compensated by material addition (synthesis) at the tip. The authors investigated the dependence of cell shape on viscosity and spatial extent of material addition; they showed that tube radius increases with viscosity and that the tip becomes more blunt as the spatial extent is increased. These results prompted a broad analysis of tip-growing cell shape from various species~\citep[plants, fungi and oomycetes;][]{campas2012}: The tip radius (radius of curvature at the tip) scales with tube radius in plants and fungi, while the tip radius appears constant in oomycetes. 

A second class of models considers growth as an incremental process whereby, at each step, the cell wall is elastically stretched by turgor pressure, and this stretched configuration is taken as the starting point of the next step; again, synthesis is assumed to keep cell wall thickness constant. \citet{goriely_biomechanical_2003} used the framework of the nonlinear elasticity theory of thin shells, which allows for large deformations of the shells. An important ingredient of their model is that the tip of the hypha is softer than the cylindrical part. Consistent with the incremental framework, growth is simulated by computing the deformation of the shell due to turgor pressure then taking the deformed shape as a new initial shape that can be deformed further. So in this model cell wall expansion is localised due to the softer tip. More recently, this model was used to study the effect of the friction between the growing tip and the external medium. This friction leads to a flattening of the tip that is consistent with experimental data \citep{goriely_mathematical_2008}. It can be shown that incremental models are mathematically equivalent to viscous models in the limit where pressure is relatively small~\citep[see for instance][]{bonazzi_symmetry_2014}; nevertheless, the interpretation of parameters differs between the two types of model. In viscous models, the viscosity (inverse of extensibility) is a proxy for the rate of cell wall remodelling under tension, but this viscosity cannot be measured directly in experiments. In elastic models, the elastic modulus (inverse of compliance) quantifies the stiffness of the wall material, which can be measured experimentally, but it does not necessarily predict how fast the cell wall expands under tension. Actually, the chemistry of the cell wall is an important ingredient that is missing from these two types of models, a limitation that applies to most of the mechanical models presented here.

In plants, root hairs and pollen tubes are quite similar to fungi with respect to the control of polar growth. Microtubules have two types of localisation in plants: cortical -- close to the plasma membrane, and endoplasmic. In root hairs and pollen tubes, microtubules, as well as actin filaments, are oriented along the tube axis and are involved in targeting the supply of new material to the tip~\citep{sieberer_microtubules_2005,gu_targeting_2013,chebli_transport_2013}. In root hairs, this organisation of microtubules depends on the cell nucleus~\citep{ambrose_microtubule_2014}. Detailed measurements in root hairs have shown that cell wall expansion occurs mainly in an annulus just behind the tip, and is isotropic there; farther from the tip, expansion becomes mostly radial and decays with distance to the tip~\citep{shaw_cell_2000}. \citet{dumais_anisotropic-viscoplastic_2006} built a mechanical model that qualitatively reproduced expansion profiles in several tip growing cells, including root hairs, using the following assumptions. The cell wall is viscoplastic: expansion occurs above a threshold in tension and then increases linearly with tension; the viscosity (inverse of extensibility) is smaller close to the tip. The thinning of the wall due to its stretching is compensated by deposition so as to keep its thickness constant. The main result is that the model accounts for quantitative measurements of cell geometry (curvatures) and wall expansion (strain rates) in root hairs only if cell walls have mechanical anisotropy. More precisely, the cell wall needs to be transversely isotropic, meaning that its properties in the direction of thickness are different from its properties in its tangent plane. Such anisotropy can be explained by the deposition of cellulose tangentially to the cell wall.

A similar pattern of expansion is observed in pollen tubes \citep{zerzour_polar_2009,hepler_control_2013}. The shape of tubes was reproduced with an incremental elastic model that included a sharp gradient of stiffness at the tip \citep{fayant_finite_2010}. The best fitting of observed shapes was achieved assuming the cell wall transversely isotropic. Interestingly, the gradient of stiffness used in the simulation is consistent with the gradients of density observed for various cell wall components, such as pectins, cellulose, and callose, that determine its mechanical properties.

As a complement to the systems studied here, we also mention trichomes --- elongated hair cells found in the aerial part of plants ---  because they appear to differ from pollen tubes. \citet{yanagisawa_growth_2015} used a viscoelastic thin-shell model of the cell wall, expanding due to turgor-generated tension. They needed to combine softer tip and mechanical anisotropy on the sides to better match experimental data.

In fungi, the local delivery of new cell wall is driven by microtubules and actin filaments. The cytoskeleton could be directly required for growth or only define the location where wall expansion takes place. Chemical treatments and mutants have demonstrated that disruption of microtubules leads to major geometrical defects in the fission yeast \citep{hagan_fission_1998}. Application of actin inhibitors can modify the dynamics of growth or completely arrest it depending on the concentrations used. Consequently, the microtubules and actin filaments must efficiently target the cell tips \citep{sawin_regulation_1998,terenna_physical_2008} to deliver the new material to the proper location. \citet{drake_model_2013} modelled the coupling between microtubule dynamics, a remodelling signal -- a protein required for cell wall remodelling, and cell wall mechanics. As in many previous studies, they considered the cell wall as effectively viscous~\citep{campas_shape_2009} and they assumed in addition that effective viscosity is reduced by the remodelling signal. They considered microtubules as growing and shrinking flexible rods attached to the nucleus. They first assumed that the level of the remodelling signal is imposed by the likelihood of contact between microtubules and cell wall, but they found that this did not enable the maintenance of cell width over many generations (as observed in living cells). Maintenance of cell width was achieved with additional assumptions: microtubules control the deposition of landmark proteins that in turn attract the remodelling signal; the remodelling signal has an intrinsic dynamics (such as reaction-diffusion) that leads to its localisation over a region of well-defined size. To summarise, \citet{drake_model_2013} built one of the first successful models for cell morphogenesis that integrates cell polarity and cell wall mechanics. It would be interesting to further probe this model by, for instance, investigating the recovery from spheroplasts (cells that became round following wall digestion) to rod-like shapes.

\citet{abenza_wall_2015} combined experiments and mechanical models to explore which cellular processes among polarity, exocytosis, or wall synthesis determine the pattern of cell wall expansion in fission yeast. They used an incremental elastic model and assumed the elastic modulus to be a function of either of the cell-end localised factors involved in the three previous cellular processes. They found that exocytosis factors better predicted the observed pattern of wall expansion. The pattern of supply of wall materials thus appears to be essential for shape, making the connection between the concepts behind mechanical models and those behind VSC-like models.

Overall, the qualitative agreement with experiments of a range of mechanical models strongly supports the notion that a softer / more compliant tip is required for tip growth. However it is yet difficult to ascertain which models are more relevant to actual cells. Further quantitative measurements of wall expansion and wall mechanics are required to make further progress.

It is generally believed that the epidermis of aerial plant tissues is under tension \citep[see][for a review]{peters_history_1996}, which would occur for instance if the epidermis is much stiffer than internal tissues \citep[as inferred in][]{beauzamy_mechanically_2015}. Consequently, a tissue like the shoot apical meristem (SAM) behaves mechanically like a pressurised shell, in which the epidermis plays the role of the shell, while inner layers corresponds to a liquid under pressure. The SAM is therefore comparable mechanically to the unicellular systems considered so far. Quantification of cell wall expansion in an Arabidopsis mutant that does not produce organs (\textit{pin-formed 1}, defective in a protein that enables efflux from cells of the phytohormone auxin) revealed higher expansion rate in an annulus that surrounds the tip \citet{kwiatkowska_surface_2004}, consistent with stiffer cell walls at the tip~\citep{milani_vivo_2011,milani_matching_2014} and reminiscent of expansion patterns in root hairs. However, this analogy is only partial because the mechanical properties of cell walls are likely anisotropic in the shoot apex (see following section). Nevertheless, spatial variations in the mechanical properties of cell walls seem to be required to establish the patterns of growth that underlie morphogenesis. For instance, the appearance of a new growth axis -- the primordium of a lateral organ such as a leaf or a flower -- on the side of the meristem is associated with a locally softer cell wall~\citep{peaucelle_pectin_2011,kierzkowski_elastic_2012}. This outgrowth requires an increase in pectin demethylesterification~\citep{peaucelle_pectin_2011}, which occurs in internal cell walls before in surface walls, and is dependent on the accumulation of auxin~\citep{braybrook_mechano-chemical_2013}. Note that this constitutes an important difference with pollen tubes, where pectin demethylesterification rigidifies the cell wall and therefore inhibits growth~\citep[see][for more details]{bosch2006}. The three-dimensional patterns in cell wall properties prompted \citet{boudon_computational_2015} to develop realistic mechanical models of tissues. Each cell wall is considered as a thin surface with elastic, plastic and viscous properties. By fine tuning the stiffness and/or viscosity of walls, \citet{boudon_computational_2015} were able to make one or several organs emerge. This work shows how heterogeneity in stiffness may generate complex shapes. Interestingly, several solutions are sometimes possible for creating a given shape. For this reason, the comparison of computational outputs with real tissues cannot be limited to shape and requires other experimental observations.

\subsection{Anisotropic diffuse growth}
The stiffness of a material is not just a number. The material can be anisotropic, i.e. it can have different values of stiffness in different directions, like for instance a fibre-reinforced material, which is harder
to stretch in the direction of the fibres. In many cases, elongation of walled cells requires such anisotropy.

In many rod-shaped bacteria, growth occurs on the cylindrical region of the cell~\citep{cava2013}. New material is inserted as small patches on the cylindrical part of the cell, a process coordinated by MreB filaments~\citep{chang_how_2014}. 

Despite the growth being distributed on their cylindrical part, bacteria grow as elongated cells. This is often thought to be caused by anisotropic reinforcement of the cell wall, either directly through a mechanical anisotropy of their material~\citep{yao_thickness_1999} or indirectly through the bacterial cytoskeleton itself. \citet{jiang_mechanical_2011} focused on the effect of MreB and built a model of elongation that reproduced shapes, divisions, and bulging in wild-type and mutant strains of \textit{E. coli}. The cell wall was considered as a continuous transversely isotropic material whose growth is driven by the transformation of intracellular energy into a mechanochemical energy, that combines the elastic energy of the cell wall and the chemical energy of new bonds. Growth is much slower than MreB dynamics, thus the mechanical effect of helical MreB filaments is averaged over time and modelled as a radial force resisting turgor pressure and yielding a preferred radius for the cell. Without this force, cells grew spherically. Whereas MreB may bend liposomes~\citep{hussain_mreb_2018}, it is unkown whether MreB is strong enough to induce curvature of the cell wall as it is synthesised.\citet{banerjee_growth_2016} extended this model to also reproduce vibrio shape, where the rod-shaped bacterium is slightly bent (e.g. \textit{Caulobacter crescentus}), by incorporating a preferred curvature of Crescentin-like proteins (with the same caveat as for MreB). It would be interesting to know whether observed cell shapes would still be retrieved by these models if forces from the cytoskeleton were replaced by anisotropy in elastic energy or in bond energy. 

Starting from a previous static model of the cell wall in Gram-negative bacteria~\citep[][see above]{huang_cell_2008}, \citet{furchtgott_mechanisms_2011} modelled wall expansion by considering the insertion of new short glycan strands into existing peptide cross-links, according to one of 3 scenarii: the first scenario --- random choice of the peptides --- leads to bulging and loss of straightness; the two other scenarii --- uniform insertion by choosing peptides inversely proportionally to their density or helical insertion according to the motion of synthase --- enables the maintenance of rod shapes.   In order to study the molecular details of cell wall remodelling, \citet{nguyen_coarse-grained_2015} later built a model on a similar, coarse-grained scale; they used more realistic mechanical parameters for the peptidoglycan network, which they inferred from molecular dynamics simulations. They achieved maintenance of rod shapes by accounting for the spatiotemporal dynamics of enzymes involved in cell wall remodelling and assuming coordination between enzyme activites. Thus, the use of different microscopic hypotheses~\citep{furchtgott_mechanisms_2011,nguyen_coarse-grained_2015} yields different macroscopic outcomes, which calls for further comparison with experimental data. In a more abstract model based on continuum mechanics, \citet{amir_dislocation_2012} represented the insertion of new material as the creation and movement of defects (dislocations in this case) along the peptidoglycan lattice. Using biological relevant values of parameters, they retrieved exponential growth in length, with a rate that is sensitive to turgor pressure. Altogether, these studies show that circumferential insertion of glycans is a key ingredient for maintenance of shape.

In bacteria, it is unclear whether the mechanical anisotropy of the cell wall is required for rod shapes, as direct experimental evidence is lacking while models with defects moving in isotropic walls~\citep{amir_dislocation_2012} may produce rod-shaped cells. In contrast, in plants, reduced cellulose content induces more isotropic growth~\citep{baskin_alignment_2001}, supporting the idea that mechanical anisotropy is required for anisotropic diffuse growth. In the multicellular context of the shoot apex, it is likely that stiffness anisotropy combines with the local softening discussed earlier. Indeed, oriented deposition of cellulose may lead to strongly anisotropic cell walls~\citep{cosgrove_growth_2005}. This would be the case in the boundary between an emerging organ and the meristem, around the tip of the shoot, and around the tip of an organ primordium~\citep{hamant_developmental_2008}, where oriented cellulose deposition is predicted based on the orientation of microtubules. The associated mechanical anisotropy would provide an additional mechanism for the elongation of the shoot or of the organ. Indeed, using a cell-based mechanical model of tissue growth, \citet{boudon_computational_2015} and \citet{sassi_auxin_2014} showed that combining changes in stiffness levels and in stiffness anisotropy ensures optimal outgrowth and better accounts for experimental observations.

\section{Feedbacks that stabilise elongated shapes}

\subsection{Sensing curvature}

An extreme case of curvature-sensing is when wall expansion is fully determined by its curvature. In tip growing cells, expansion is maximal at the tip, where the wall curvature is the highest. \citet{goriely_growth_2005} and \citet{jaffar_basic_2013} built geometrical models of tip-growing cells in which local wall expansion (and accordingly material supply) is an increasing function of local curvature (Gauss curvature in 3D models). \citet{jaffar_basic_2013} used two fitting parameters to reproduce the geometry of cells of many organisms (plants, fungi, actinobacteria), though the biological relevance of these parameters is unclear. Such models show that curvature-based expansion is sufficient to account for the stable form of tip-growing cells.

There is evidence for curvature-sensing in bacteria. \citet{ursell_rod-like_2014} and \citet{billings_novo_2014} found that MreB localises preferentially to regions where the wall has negative curvature; the spatiotemporal correlation of expansion and MreB indicates that MreB precedes expansion. This causal link between curvature and MreB was confirmed by perturbation experiments. Altogether, experimental data suggest that MreB relocalises to regions of negative curvature, inducing more expansion and reverting the local geometry to cylindrical. \citet{ursell_rod-like_2014} used the model for glycan strand insertion \citep[][see above]{furchtgott_mechanisms_2011} to test this mechanism in \textit{E. coli}. They assumed that insertion occurs preferentially at regions of low curvature and found that this stabilised growth of a rod-shaped cell and enabled the recovery from an initially bent shape. \citet{hussain_mreb_2018} gave further experimental support to these conclusions by manipulating shapes of \textit{Bacillus subtilis} by chemical treatments or by confinement in channels. However, \citet{wong_mechanical_2017} combined experimental and theoretical approaches to show that the observed local enrichment in MreB in deformed bacteria is not sufficient to explain the recovery of a straight shape.

\subsection{Sensing forces}
Bacteria are known to sense their mechanical state through channels that are sensitive to membrane tension. In \textit{E. coli}, \citet{wong_mechanical_2017} proposed that strain-activated growth could qualitatively explain the response and recovery of experimentally bent bacteria. Interestingly, although the biochemical implementation of this mechanism is unclear, it was suggested to work jointly with MreB-mediated regulation (as discussed above). 

In fission yeast, spores grow roughly spherically, until an outgrowth initiates the rod-like shape of vegetative cells. The associated transition from unstable to stable polarity is triggered mechanically by the rupture of the outer wall of the spore, which is a very stiff thin layer that surrounds the vegetative-like wall~\citep{bonazzi_symmetry_2014}. Interestingly, the ratio between the volume at the transition and the initial volume of the spore is constant, despite a large variability in initial volume. \citet{bonazzi_symmetry_2014} modelled the outer wall as an elastic stiff shell (that may rupture) and considered the vegetative wall either as elastic or as viscoplastic --- the viscoplastic region is where growth occurs and it corresponds to the location of the polarisome (the ensemble of polarly localised proteins). The polarisome was assumed to move randomly, mimicking experimental observations. When it is intact, the outer shell prevents outgrowth and keeps the spore roughly spherical. The tension in the shell increases, up to a threshold over which its local rupture initiates the outgrowth. The theoretical results thus show that stress-sensing via mechanical rupture enables the outgrowth to occur when the ratio of spore volume to initial volume has reached a well-defined threshold. 

\begin{figure}
\begin{centering}
\includegraphics[scale=0.3]{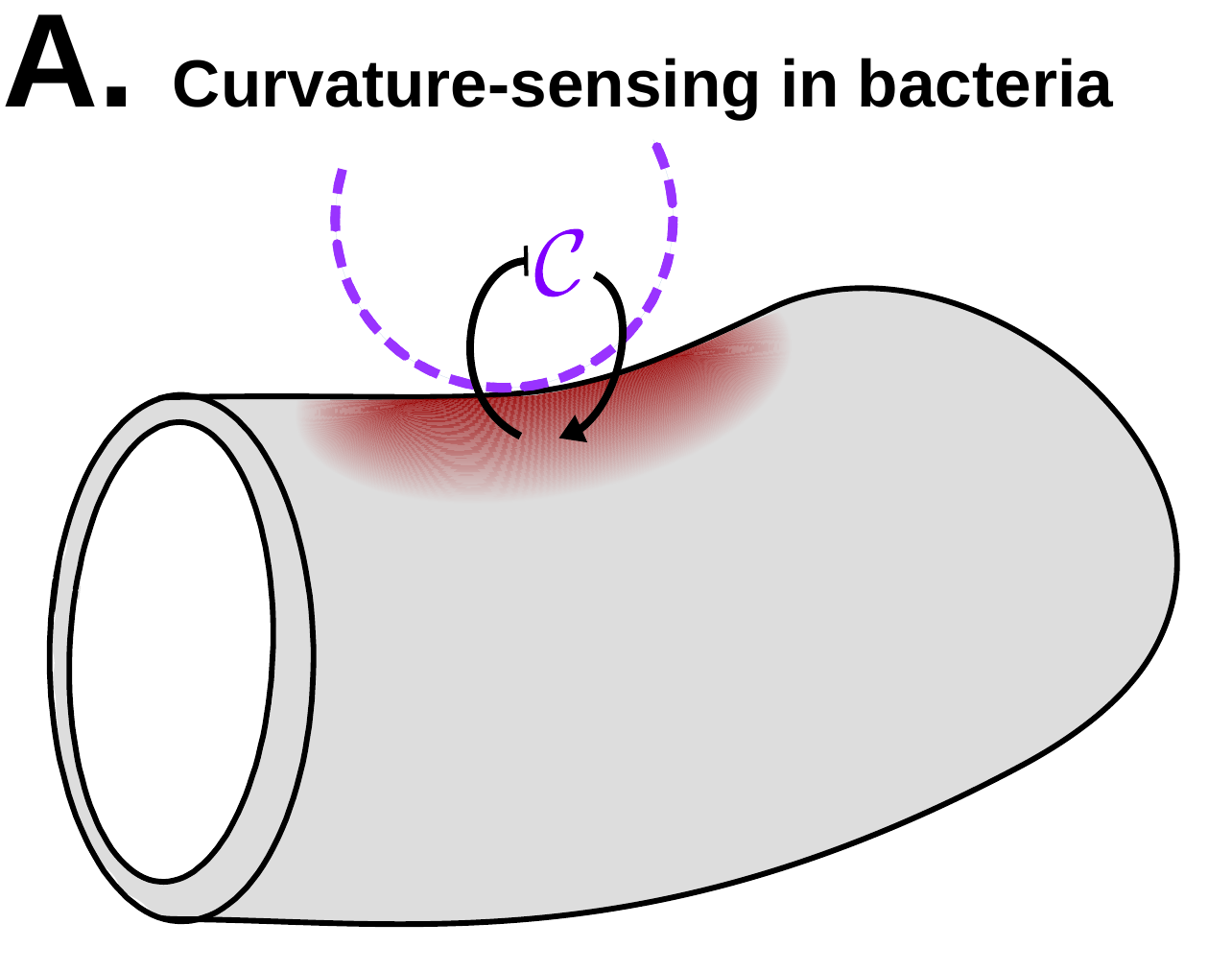}
\par\end{centering}

\begin{centering}
\includegraphics[scale=0.3]{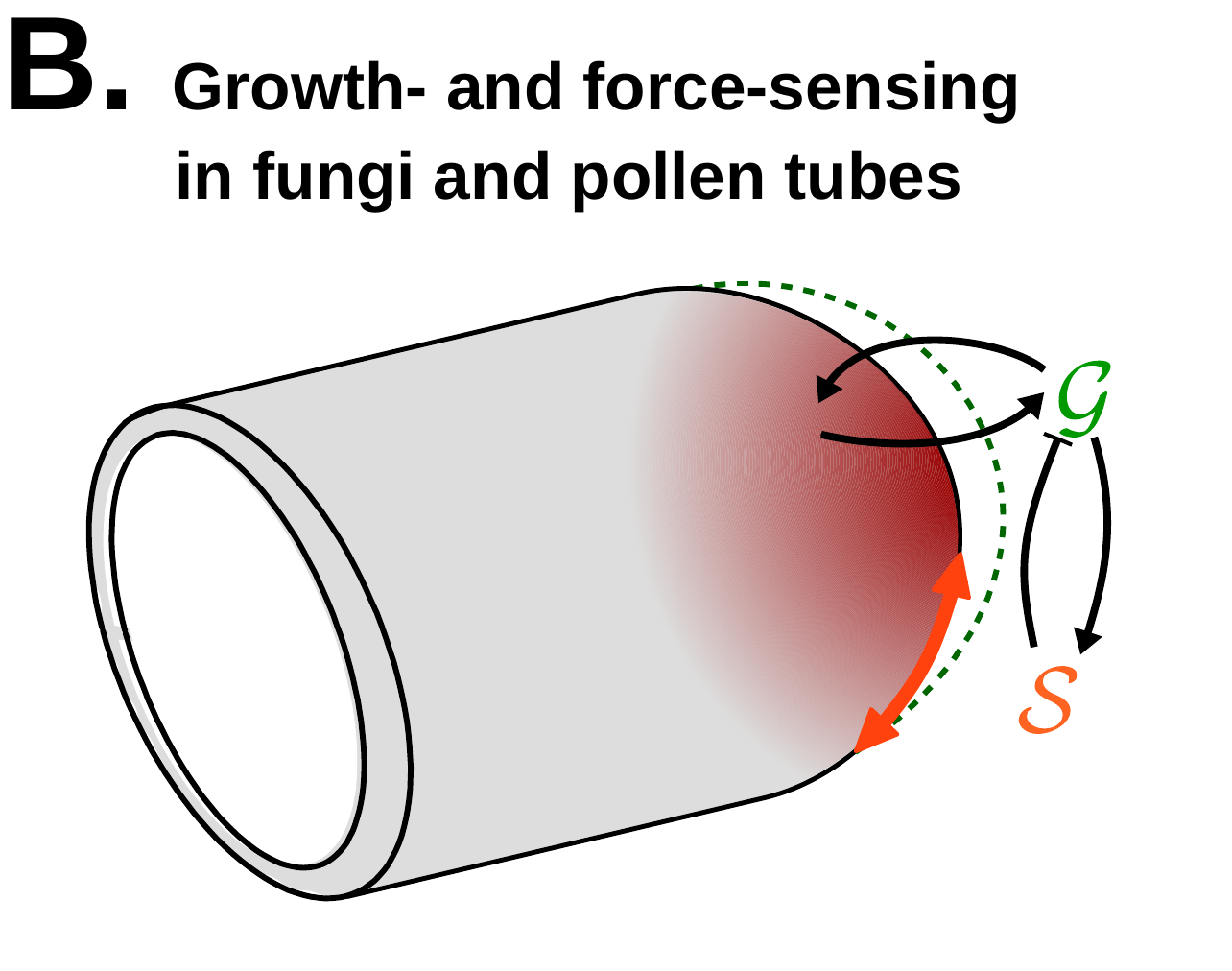}
\par\end{centering}

\begin{centering}
\includegraphics[scale=0.3]{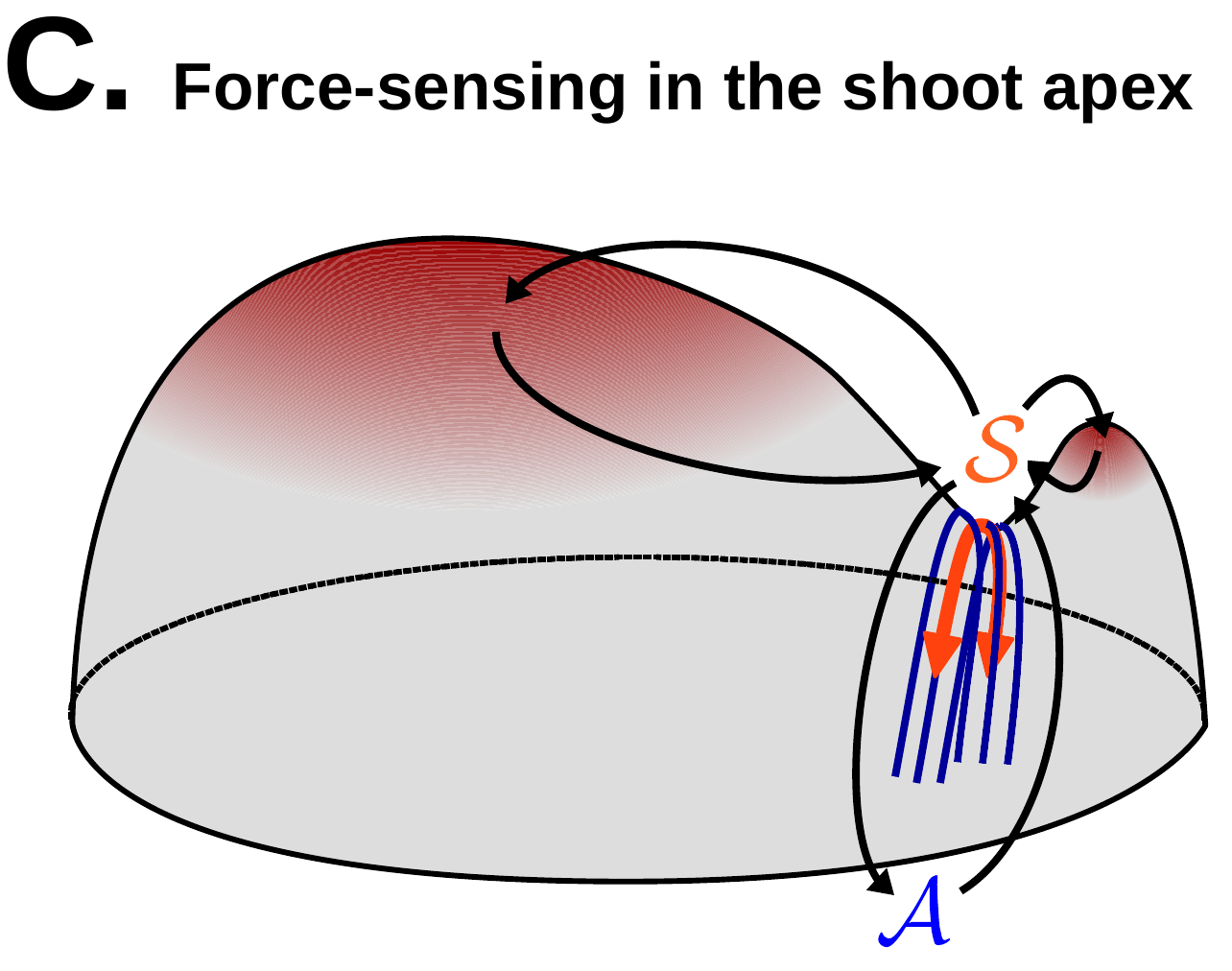}
\par\end{centering}

\caption{\textbf{Feedbacks that stabilise elongation.} (A) In \textit{E. coli}, the insertion of new cell wall is increased in the region of negative curvature. This feedback may stabilise rod shape and
enable recovery from initially curved shapes \citep{ursell_rod-like_2014,billings_novo_2014,hussain_mreb_2018}. (B) In pollen tubes and fission yeast, surface expansion feeds back on material supply. In pollen tubes this feedback may lead to oscillatory tip growth \citep{rojas_chemically_2011}. In fission yeast, this feedback occurs through the position of the polar cap (where polarity proteins are localised). It leads to the random shuffling of polar cap in spores. After the rupture of the outer spore wall, the feedback promotes tip-growth \citep{bonazzi_symmetry_2014}. (C) In the plant shoot apex, two loops involving mechanosensation are coupled. By enhancing transport of the phytohormone auxin, mechanical strain and stress focus growth at the tip of the organ. Mechanical stress may also increase mechanical anisotropy of the cell wall via deposition of cellulose oriented by the response of microtubules to mechanical stress~\citep{hamant_developmental_2008,heisler_alignment_2010}.}
\end{figure}

As mentioned above, morphogenesis at the shoot apex relies both on local softening and mechanical anisotropy. The two mechanisms are tuned by a feedback from mechanics. The local softening that initiates organ emergence is triggered by the local accumulation of the phytohormone auxin~\citep{sassi_auxin_2013}. Auxin patterns are determined by the polarity of PIN FORMED1 (PIN1), a membrane-addressed protein that facilitates auxin efflux. Three hypotheses have been proposed for the determination of auxin polarity based on experimental observations~\citep{abley_auxin_2013,sassi_auxin_2013}: flux of auxin, gradient of auxin, and intrinsic property of the cell that may be oriented by external cues. We here focus on the second hypothesis, because it has been related to mechanical signals. According to this hypothesis, PIN1 polarity in a cell is determined by auxin concentration in neighbouring cells so that PIN1 is polarised towards the cell with higher concentration. This so-called ``up the gradient'' model can reproduce observed auxin patterns~\citep{jonsson_auxin-driven_2006,smith_plausible_2006}, though it raises questions about how cells might sense auxin concentration in neighbouring cells.  More recent experimental evidence suggests that PIN1 proteins indirectly sense the mechanical status of cell walls~\citep{heisler_alignment_2010,nakayama_mechanical_2012,braybrook_mechano-chemical_2013}. More precisely, PIN1 in a cell would be polarised towards the cell wall with the highest mechanical stress/strain (which of strain or stress is sensed is still unclear), which is shared with the cell with highest auxin concentration due to the induced softening of its  walls. This chemomechanical model has been implemented using the finite element method for the mechanics coupled with a system of differential equation for the auxin dynamics \citep{heisler_alignment_2010}. It is able to generate patterns of auxin accumulation and to reproduce the radial PIN1 reorientation observed around a cell ablation.

In many plant tissues, mechanical signals also regulate the orientation of cellulose microfibrils~\citep{castle1937,green1966,preston1988,wasteneys1987,wasteneys1989,williamson1990,fischer1997,hejnowicz2000,hamant_developmental_2008,jacques2013,sampathkumar2014}. Indeed, cellulose is synthesised following the orientation of cortical  microtubules~\citep{baskin_alignment_2001,bringmann2012}. Additionally, microtubules orient along the direction of maximal mechanical tension~\citep{wasteneys1987,wasteneys1989,williamson1990,fischer1997,hejnowicz2000,hamant_developmental_2008,jacques2013,sampathkumar2014}. Consequently, the preferential orientation of cellulose microfibrils reinforces the cell wall in the direction of mechanical stress \citep{landrein_how_2013}. Several theoretical studies investigated the consequences of this feedback loop between mechanical stress and cell wall anisotropy. \citet{hamant_developmental_2008} modelled the shoot apical meristem as an elastic surface in 3D, thus only accounting for the epidermal cell layer. They used a vertex model, meaning that they only considered cell walls orthogonal to the surface of epidermis, represented as 1-dimensional springs. The stiffnesses of these springs increases as they are more parallel to the local mechanical tension, mimicking the orientation of the microtubules and their feedback on the mechanical properties of cell walls. Growth is driven by the turgor pressure of internal tissues. Above a stress threshold, the cell walls yield and thus deform plastically. By initiating the emergence of an organ via the local softening of a group of cells, the cellulose reorientation leads to a circumferential pattern around the organ, reinforcing the boundary with the apical dome and thus making the symmetry breaking more effective. \citet{bozorg_stress_2014} obtained similar results using a continuous model for the epidermal layer. They further showed that using mechanical strain instead of stress as a directional cue is not sufficient to account for experimental observations.

\subsection{Sensing growth rate}
In fission yeast, an additional result from the mechanical model discussed above~\citep{bonazzi_symmetry_2014} is that a positive feedback between growth and polarity can explain the stabilisation during spore outgrowth.  \citet{bonazzi_symmetry_2014} considered three possible cues that bias the random motion of the polarisome (polarly localised proteins): curvature of the cell surface, mechanical stress in the surface, and expansion rate. Only the last cue led to stable cylindrical shapes, which was supported by further experiments in which wall expansion was manipulated.

\citet{rojas_chemically_2011} developed a model coupling the deposition of new material and the mechanics of the cell wall. They reproduced the morphologies of pollen tubes and were able to explain growth oscillations that are observed in rapidly growing tubes. In this model, the rate of deposition of wall material decreases with the velocity of cell tip, making a link between exocytosis and growth rate. Consequently, the cell may either grow at constant rate or oscillate between phases of high deposition and slow elongation and phases of low deposition and fast elongation. This results in either cylindrical or pearled pollen tubes, respectively. At least two hypotheses could account for such negative feedback. A `passive' hypothesis is that when growth velocity increases, exocytosis becomes relatively too slow to provide materials to the growing tip. An `active' hypothesis is that a high rate cell wall of expansion may lead to higher membrane tension (due to limited membrane supply), leading to the opening of mechanosensitive channels and the entry of cytosolic calcium that would downregulate the polymerisation of the actin cytoskeleton and thus the delivery of cell wall material~\citep{kroeger_model_2008,yan_calcium_2009}. 

\citet{rojas_homeostatic_2017} combined experiments in \textit{Bacillus subtilis} and a non-spatialised dynamical model to also propose that enhanced expansion would induce high membrane tension. Mechanosensing would prevent over-expansion of the wall by reducing the supply of wall precursors when membrane tension is too large. It would be interesting to know whether this mechanisms is involved in regulating cell shape. Overall, the three studies discussed use models to suggest that the rate of cell wall expansion is sensed, though the mechanisms behind are still to be identified.

\section{Conclusions}

The generation of anisotropic shapes in walled cells relies mainly on two strategies. Many cells, such as hyphae, yeasts, root hairs or pollen tubes grow directionally via the supply of new material to the cell wall at a precise and restricted location \citep{tindemans_diffusive_2006,drake_model_2013}. Several models for tip growth have been implemented and are able to reproduce most of observed shapes. These computational approaches may give some information about the mechanics of the cell wall \citep{dumais_anisotropic-viscoplastic_2006,fayant_finite_2010} or about its behaviour with respect to  perturbations \citep{goriely_mathematical_2008}. However, the range of hypotheses used in these models makes it difficult to know which ones are more relevant to experiments. Progress should stem from more quantitative experiments and models, and from the study of perturbations in experiments and in models. Rod-shaped bacteria grow diffusely, requiring the mechanical reinforcement of their sides. This reinforcement is likely achieved thanks to the circumferential insertion of glycan strands \citep{huang_cell_2008,furchtgott_mechanisms_2011}. On a multicellular scale, plants combine those tip growth and anisotropic diffuse growth to initiate organs morphogenesis. Theoretical models indicate that the two mechanisms are necessary to induce the massive shape changes required for organogenesis. The major actors of this mechanical control of morphogenesis, each corresponding to one of the strategies discussed here, are the phytohormone auxin, which is involved in the softening of the plant cell wall~\citep{reinhardt_auxin_2000}, and cellulose, a stiff polymer whose oriented deposition leads to stiffness anisotropy~\citep{baskin_alignment_2001}.

Several feedbacks have been identified that may contribute to the maintenance of rod-shapes. Bacteria, thanks to a simple mechanism of curvature-sensing based on the MreB protein, are able to maintain and even to generate \textit{de novo} cylindrical shapes \citep{billings_novo_2014,ursell_rod-like_2014,wong_mechanical_2017}. In fission yeast, a precise volume doubling between the germination and the outgrowth is granted by the mechanical rupture of its protective shell \citep{bonazzi_symmetry_2014}. Force-sensing is also involved in  organogenesis in the plant shoot apex. Mechanical stress, auxin transport, auxin-induced softening and cellulose anisotropy feed back on each other in complex loops that may be required for the robustness of organogenesis~\citep{hamant_developmental_2008,heisler_alignment_2010}. Finally, growth-sensing explains both the random movement and the stabilisation of polarity before and after the triggering of the outgrowth in fission yeast~\citep{bonazzi_symmetry_2014}. Growth-sensing might also be relevant for the oscillatory growth in pollen tubes \citep{rojas_chemically_2011}. All these studies show how theoretical approaches may help unravelling the complex feedbacks that underly organismal growth and robustness of morphogenesis. Simulations of these models enable testing alternative hypotheses that can be difficult to differentiate experimentally or may lead to the identification of key experiments.

The molecular actors behind many of these feedbacks are unknown. Curvature-sensing in bacteria could be due to a membrane curvature-dependent binding energy of the protein complex that includes MreB. Negative curvature and MreB localisation could also be driven by a common signal such as proteins involved in cell wall synthesis~\citep{billings_novo_2014}. In fission yeast, mechanisms similar to the oscillatory growth of pollen tubes could explain the stabilisation of polarity by surface expansion~\citep{yan_calcium_2009,rojas_chemically_2011}. Alternatively, polarity could be diluted and destabilised in the absence of sufficient growth \citep{layton_modeling_2011}. Finally, growth could be involved in the monitoring of cellular dimensions by intracellular gradients \citep{howard_how_2012}. In the context of the shoot apex, feedback mechanisms are still poorly understood and we may only speculate. Stretching of the cell membrane or of the cell wall could activate ion channels or modify the conformation of wall-bound proteins, triggering pathways that impact on microtubules or on PIN1 \citep{landrein_how_2013}. In the case of auxin transport, an alternative hypothesis is the activation of exocytosis and inhibition of endocytosis by the tension in the cell membrane \citep{hamill_molecular_2001}. All these hypotheses lack evidence, but new insights can be expected with progress in cellular and developmental biology, together with physically-based models of the associated processes.

We tried here to highlight concepts and generic mechanisms that hold across kingdoms. Future directions might stem from enhanced cross-fertilisation between approaches and concepts developed in the context of specific systems. For instance, detailed mechanical models of the growing bacterial cell wall could provide inspiration for the more complex and less organised cell walls of fungi, oomycetes, and plants. Conversely, continuous models developed for plants, fungi, and oomycetes could be used to test coarse hypotheses in bacterial morphogenesis, before dealing with more involved molecular details. More generally, modelling morphogenesis at multiple scales, with models assembled as a Russian doll to make links between successive scales or levels, would allow to deal with a suite of simple models than can be falsified separately and assembled to address more elaborate questions.

\section*{Acknowledgements}
AB is supported by Institut Universitaire de France. AB would like to thank the Isaac Newton Institute for Mathematical Sciences for support and hospitality during the programme Growth form and self-organisation when this paper was finalised. This work was partially supported by EPSRC grant number EP/K032208/1 and by a grant from the Simons Foundation. 

%% The Appendices part is started with the command \appendix;
%% appendix sections are then done as normal sections
%% \appendix

%% \section{}
%% \label{}

%% If you have bibdatabase file and want bibtex to generate the
%% bibitems, please use
%%
%%  \bibliographystyle{elsarticle-harv} 
%%  \bibliography{<your bibdatabase>}

%% else use the following coding to input the bibitems directly in the
%% TeX file.

%\begin{thebibliography}{00}

%% \bibitem[Author(year)]{label}
%% Text of bibliographic item

%\bibitem[ ()]{}

%\end{thebibliography}

\end{document}